\documentclass{article}

\usepackage{fullpage}
\usepackage{amssymb, amsmath, amsthm}
\usepackage{graphicx}
\usepackage{xspace}
\usepackage{todonotes}
\usepackage[all=normal,bibliography=tight]{savetrees}
\usepackage{enumerate}

\newtheorem{theorem}{Theorem}
\newtheorem{lemma}[theorem]{Lemma}

\newtheorem{definition}[theorem]{Definition}
\newtheorem{corollary}[theorem]{Corollary}
\newtheorem{claim}[theorem]{Claim}

\def\cqedsymbol{\ifmmode$\lrcorner$\else{\unskip\nobreak\hfil
\penalty50\hskip1em\null\nobreak\hfil$\lrcorner$
\parfillskip=0pt\finalhyphendemerits=0\endgraf}\fi} 

\newcommand{\cqed}{\renewcommand{\qed}{\cqedsymbol}}

\newcommand{\executeiffilenewer}[3]{%
\ifnum\pdfstrcmp{\pdffilemoddate{#1}}%
{\pdffilemoddate{#2}}>0%
{\immediate\write18{#3}}\fi%
} 

\newcommand{\Oh}{\ensuremath{\mathcal{O}}}

\newcommand{\pdeg}{\delta}
\newcommand{\pconst}{C}
\newcommand{\Ball}{B}
\newcommand{\DBall}{\partial B}
\DeclareMathOperator{\dist}{dist}

\title{Subexponential parameterized algorithms for graphs of polynomial growth\thanks{%
The research of D. Marx leading to these results has received funding from the European Research Council under the European Union's Seventh Framework Programme (FP/2007-2013) / ERC Grant Agreement no.~280152.
The research of M. Pilipczuk is supported by Polish National Science Centre grant UMO-2013/09/B/ST6/03136.
Part of the research has been done when the authors were participating in the ``Fine-grained complexity and algorithm design'' program
at the Simons Institute for Theory of Computing in Berkeley.}}

\date{}

\author{D\'{a}niel Marx\thanks{Institute for Computer Science and Control, Hungarian Academy of Sciences (MTA SZTAKI), Hungary, \texttt{dmarx@cs.bme.hu}. }
  \and Marcin Pilipczuk\thanks{Institute of Informatics, University of Warsaw, Poland, \texttt{marcin.pilipczuk@mimuw.edu.pl}.}}

\begin{document}

\maketitle

\begin{abstract}
  We show that for a number of parameterized problems for which only
  $2^{\Oh(k)} n^{\Oh(1)}$ time algorithms are known on general graphs,
  subexponential parameterized algorithms with running time
  $2^{\Oh(k^{1-\frac{1}{1+\pdeg}} \log^2 k)} n^{\Oh(1)}$ are possible
  for graphs of polynomial growth with growth rate (degree) $\pdeg$,
  that is, if we assume that every ball of radius $r$ contains only
  $\Oh(r^\pdeg)$ vertices. The algorithms use the technique of {\em
    low-treewidth pattern covering,} introduced by Fomin et
  al.~\cite{patcov-focs} for planar graphs; here we show how this strategy can be made to work for graphs with polynomial growth.

 Formally, we prove that,
  given a graph $G$ of polynomial growth with growth rate $\pdeg$ and
  an integer $k$, one can in randomized polynomial time find a subset
  $A \subseteq V(G)$ such that on one hand the treewidth of $G[A]$ is
  $\Oh(k^{1-\frac{1}{1+\pdeg}} \log k)$, and on the other hand for
  every set $X \subseteq V(G)$ of size at most $k$, the probability
  that $X \subseteq A$ is $2^{-\Oh(k^{1-\frac{1}{1+\pdeg}} \log^2
    k)}$.  Together with standard dynamic programming techniques on
  graphs of bounded treewidth, this statement gives subexponential
  parameterized algorithms for a number of subgraph search problems,
  such as \textsc{Long Path} or \textsc{Steiner Tree}, in graphs of
  polynomial growth.

  We complement the algorithm with an almost tight lower bound for
  \textsc{Long Path}: unless the Exponential Time Hypothesis fails, no
  parameterized algorithm with running time
  $2^{k^{1-\frac{1}{\pdeg}-\varepsilon}}n^{\Oh(1)}$ is possible for
  any $\varepsilon > 0$ and an integer $\pdeg \geq 3$.
\end{abstract}

\section{Introduction}
In recent years, research on parameterized algorithms had a strong focus on understanding the optimal form of dependence on the parameter $k$ in the running time $f(k)n^{\Oh(1)}$ of parameterized algorithms. For many of the classic algorithmic problems on graphs, algorithms with running time $2^{\Oh(k)}n^{O(1)}$ exist, and we know that this form of running time is best possible, assuming the Exponential-Time Hypothesis (ETH) \cite{eth,thebook,eth-surveye}. This means that we have an essentially tight understanding of these problems when considering graphs in their full generality, but does not rule out the possibility of improved algorithms when restricted to some class of graphs. Indeed, many of these problems become significantly easier on certain important graph classes. The most well-studied form of this improvement is the so-called ``square root phenomenon'' on planar graphs (and some if its generalizations): there is a large number of parameterized problems that admit $2^{O(\sqrt{k}\cdot\textup{polylog} k)}n^{O(1)}$ time algorithms on planar graphs \cite{DBLP:journals/siamdm/DemaineFHT04,DBLP:journals/talg/DemaineFHT05,DemaineFHT05,DBLP:journals/cj/DemaineH08,DBLP:journals/combinatorica/DemaineH08,DBLP:conf/gd/DemaineH04,DBLP:journals/csr/DornFT08,DornPBF10,DBLP:journals/siamcomp/FominT06,DBLP:conf/esa/Thilikos11,DBLP:journals/ipl/FominLRS11,DBLP:conf/stacs/DornFLRS10,DBLP:conf/icalp/KleinM12,DBLP:conf/soda/KleinM14,DBLP:conf/soda/ChitnisHM14,DBLP:conf/stacs/PilipczukPSL13,DBLP:conf/focs/PilipczukPSL14}. Many of these positive results can be explained by the theory of bidimensionality \cite{DemaineFHT05} and explicity or implicitly relies on the relation between treewidth and grid minors.

Very recently, a superset of the present authors showed a new
technique to obtain subexponential algorithms in planar graphs for
problems related to the \textsc{Subgraph Isomorphism}
problem~\cite{patcov-arxiv,patcov-focs}, such as the \textsc{Long
  Path} problem of finding a simple path of length $k$ in the input
graph.  The approach of~\cite{patcov-arxiv,patcov-focs} can be
summarized as follows: a randomized polynomial-time algorithm is
showed that, given a planar graph $G$ and an integer $k$, selects a
random induced subgraph of treewidth sublinear in $k$ in such a
manner that, for every connected $k$-vertex subgraph $H$ of $G$, the
probability that $H$ survives in the selected subgraph is
inversely-subexponential in $k$. Such a statement, dubbed
\emph{low-treewidth pattern covering}, together with standard dynamic
programming techniques on graphs of bounded treewidth, gives
subexponential algorithms for a much wider range of \textsc{Subgraph
  Isomorphism}-type problems than bidimensionality; for example, while
bidimensionality provides a subexponential algorithm for \textsc{Long
  Path} in undirected graphs, it seems that the new approach
of~\cite{patcov-arxiv,patcov-focs} is needed for directed graphs.

The proof of the low treewidth pattern covering theorem of~\cite{patcov-arxiv,patcov-focs} involves a number of different partitioning techniques in planar graphs.
In this work, we take one of this technique --- called \emph{clustering procedure}, based on the metric decomposition tool of Linial and Saks~\cite{LinialS93} and the recursive decomposition used in the construction
of Bartal's HSTs~\cite{Bartal98} --- and observe that it is perfectly suited to tackle the so-called \emph{graphs of polynomial growth}.

To explain this concept formally, let us introduce some notation. 
All graphs in this paper are unweighted, and the distance function $\dist_G(u,v)$
measures the minimum possible number of edges on a path from $u$ to $v$ in $G$.
For a graph $G$, integer $r$, and vertex $v \in V(G)$ by $\Ball_G(v, r)$ we
denote the set of vertices $w \in V(G)$ that are within distance \emph{less than $r$}
from $v$ in $G$, $\Ball_G(v, r) = \{w \in V(G): \dist_G(v, w) < r\}$, 
while by $\DBall_G(v, r)$ we denote the set of vertices within distance
\emph{exactly $r$}, that is, $\DBall_G(v, r) = \{w \in V(G): \dist_G(v, w) = r\}$.
We omit the subscript if the graph is clear from the context.

\begin{definition}[polynomial growth, \cite{BlondelJKS13}]
We say that a graph $G$ (or a graph class $\mathcal{G}$) has \emph{polynomial growth} of degree (growth rate) $\pdeg$ if there exists a universal constant $\pconst$ such that for (every graph $G \in \mathcal{G}$ and)
every radius $r$ and every vertex $v \in V(G)$ we have
$$|\Ball(v, r)| \leq \pconst \cdot r^\pdeg.$$
\end{definition}
The algorithmic consequences (and some of its variants) of this definition have been studied in the literature in various contexts (see, for example, \cite{DBLP:conf/spaa/AbrahamM05,DBLP:conf/infocom/GummadiJSS09,BlondelJKS13,AbrahamGGM06}).
A standard example of a graph of polynomial growth with degree $\pdeg$ is a $\pdeg$-dimensional grid. 
Graph classes of polynomial growth include graphs of bounded doubling dimension (with unit-weight edges), a popular
assumption restricting the growth of a metric space in approximation algorithms or
routing in networks (cf. the thesis~\cite{chan-thesis} of Chan or~\cite{AbrahamGGM06} and references therein).

The clustering procedure, or metric decomposition tool of~\cite{LinialS93}, can be described as follows. As long as the analysed graph $G$ is not empty, we carve out a new cluster as follows.
We pick any vertex $v \in V(G)$ as a center of the new cluster, and set its radius $r := 1$. Iteratively, with some chosed probablity $p$, we accept the current radius, and with the remaining
probability $1-p$ we increase $r$ by one and repeat. That is, we choose $r$ with geometric distribution with success probability $p$. Once a radius $r$ is accepted, we set $\Ball_G(v,r)$ as a
new cluster, and delete $\Ball_G(v,r) \cup \DBall_G(v,r)$ from $G$. In this manner, $\Ball_G(v,r)$ is carved out as a separated cluster, at the cost of sacrificing $\DBall_G(v,r)$.
A typical usage would be as follows:
If one choose $p$ of the order of $k^{-1}$, then a simple analysis shows that every cluster has radius $\Oh(k \log n)$ w.h.p., while a fixed set $X \subseteq V(G)$ of size $k$ is fully retained in \
the union of clusters with constant probability.
By a careful two-step application of the clustering procedure, we show the following low treewidth pattern covering statement for graphs of polynomial growth.

\begin{theorem}\label{thm:alg}
For every graph class $\mathcal{G}$ of polynomial growth with growth rate $\pdeg$,
there exists a polynomial-time randomized algorithm that, given a graph $G \in \mathcal{G}$
and an integer $k$, outputs a subset $A \subseteq V(G)$ with the following properties:
\begin{enumerate}
\item treedepth of $G[A]$ is $\Oh(k^{1 - \frac{1}{1+\pdeg}} \log k)$;
\item for every set $X \subseteq V(G)$ of size at most $k$, the probability that
$X \subseteq A$ is $2^{-\Oh(k^{1-\frac{1}{1+\pdeg}} \log^2 k)}$.
\end{enumerate}
\end{theorem}
Note that Theorem~\ref{thm:alg} uses the notion of \emph{treedepth}, a much more restrictive
graph measure than treewidth (cf.~\cite{sparsity}),
that in particular implies the same treewidth bound.
Thus, together with standard dynamic programming techniques on graphs of bounded treewidth, Theorem~\ref{thm:alg} gives
the following.
\begin{corollary}
There exist randomized parameterized algorithms with running time bound $2^{\Oh(k^{1-\frac{1}{1+\pdeg}} \log^2 k)} n^{\Oh(1)}$
for \textsc{Long Path}, \textsc{Vertex Cover Local Search}, and \textsc{Steiner Tree} parameterized by the size of the solution tree,
when restricted to a graph class of polynomial growth with growth rate $\pdeg$.
\end{corollary}
We refer to the introduction of~\cite{patcov-arxiv} for bigger discussion on applications of low treewidth pattern covering statements.

We complement the algorithmic statement of~Theorem~\ref{thm:alg} with the following lower bound.
\begin{theorem}\label{thm:lb}
If there exists an integer $\pdeg \geq 3$, a real $\varepsilon > 0$,
and an algorithm that decides if a given subgraph of a $\pdeg$-dimensional
grid of side length $n$ contains a Hamiltonian path in time $2^{\Oh(n^{\pdeg-1-\varepsilon})}$,
then the ETH fails.
\end{theorem}
Since a subgraph of a $\pdeg$-dimensional grid of side length $n$ has polynomial growth with degree at most $\pdeg$ and at most $n^{\pdeg}$ vertices, Theorem~\ref{thm:lb} shows that,
unless the ETH fails, one cannot hope for a better term than $k^{1-\frac{1}{\pdeg}}$ in the low treewidth pattern covering statement as in Theorem~\ref{thm:alg}.

\section{Upper bound: proof of Theorem~\ref{thm:alg}}\label{sec:upper}

In this section we prove Theorem \ref{thm:alg}. Without loss of generality, we assume $k \geq 4$.

The algorithm works in two steps. In the first one, the goal is to chop the graph into components
of radius $\Oh(k \log k)$, which --- by the polynomial growth property --- are of polynomial size.
Then, in the second phase, we consider every component independently, sparsifying it further.
These two steps are described in the subsequent two section.

\subsection{Chopping the graph into parts of polynomial size}

The goal of the first step is to delete a number of vertices from the graph so that on one hand every connected component of $G$ has radius $\Oh(k \log k)$,
and on the other hand the probability of deleting a vertex from an unknown pattern $X \subseteq V(G)$ of size at most $k$ is small. 
Formally, we show the following lemma.

\begin{lemma}\label{lem:alg1}
Let $\mathcal{G}$ be as in Theorem~\ref{thm:alg}.
There exists a constant $c_r > 0$ and a polynomial-time randomized algorithm that, given a graph $G \in \mathcal{G}$
and positive integer $k \geq 4$, outputs a subset $A \subseteq V(G)$ such that
\begin{enumerate}
\item every connected component of $G[A]$ is of radius at most $c_r k \log k$;
\item for every set $X \subseteq V(G)$ of size at most $k$, the probability that
$X \subseteq A$ is at least $17/256$.
\end{enumerate}
\end{lemma}

\begin{proof}
For a constant $c_r > 0$ to be fixed later, we perform the following iterative process.
We start with $G_0 := G$ and $B_0 := \emptyset$. In $i$-th iteration ($i=1,2,3,\ldots$), we consider the graph $G_{i-1}$.
If the graph $G_{i-1}$ is empty, we stop. Otherwise, we pick an arbitrary vertex $v_i \in V(G_{i-1})$ and pick a radius $r_i$
according to the geometric distribution with success probability $1/k$, capped at value $R := c_r k \log k$ (i.e., if the choice of the radius is greater than $R$, we set $r_i := R$).
For further analysis, we would like to look at the choice of the radius $r_i$ as the following iterative process: we start with $r_i=1$ and iteratively
accept the current radius with probability $1/k$ or increase it by one and repeat with probability $1-1/k$, stopping unconditionally at radius $R$.
Given $v_i$ and $r_i$, we set $A_i := A_{i-1} \cup \Ball_{G_{i-1}}(v_i, r_i)$
and $G_i := G_{i-1} - (\Ball_{G_{i-1}}(v_i, r_i) \cup \DBall_{G_{i-1}}(v_i, r_i))$. 
That is, we remove from $G_i$ all vertices within distance at most $r_i$ from $v_i$, while retaining in $A_i$ only those that are within distance less than $r_i$.

Clearly, as we remove a vertex from $G_i$ at every step, the process stops after at most $|V(G)|$ steps.
Let $i_0$ be the last index of the interation. Consider the graph $G' := G[A_{i_0}]$.
Recall that in the $i$-th step we put $\Ball_{G_{i-1}}(v_i,r_i)$ into $A_i$, but remove not only $\Ball_{G_{i-1}}(v_i,r_i)$
from $G_{i-1}$ but also $\DBall_{G_{i-1}}(v_i,r_i) = N_{G_{i-1}}(\Ball_{G_{i-1}}(v_i,r_i))$. Consequently, 
the vertex sets of the connected components of $G'$ are exactly sets $\Ball_{G_{i-1}}(v_i, r_i)$ for $1 \leq i \leq i_0$.
Since the radii $r_i$ are capped at value $R = c_r k \log k$, every connected component of $G'$ has radius at most $R$.

We now claim the following.
\begin{claim}\label{cl:chop-hit}
For every $X \subseteq V(G)$ of size at most $k$, the probability that $X \subseteq V(G')$ is at least $17/256$.
\end{claim}
\begin{proof}
Fix $X \subseteq V(G)$ of size at most $k$. Note that $X \not\subseteq V(G')$ only if at some iteration $i$, some vertex $x \in X$ is exactly within distance $r_i$ from $v_i$ in the graph $G_{i-1}$.
We now bound the probability that this happens, split into two subcases: either $r_i = R$ or $r_i < R$.

\textbf{Case 1: hitting a vertex within distance $r_i = R$.}
Let $Y = \bigcup_{x \in X} \Ball_G(x, R+1)$. Note that if $x \in X$ is exactly within distance $r_i \leq R$ from $v_i$ in the graph $G_{i-1}$, then necessarily $v_i \in Y$.
On the other hand, by the polynomial growth property,
   $$|Y| \leq k \cdot \pconst \cdot (R+1)^\pdeg = \pconst k (c_r k \log k + 1)^\pdeg = \Oh(k^{\pdeg + 1} \log^\pdeg k).$$

We consider ourselves \emph{lucky} if whenever $v_i \in Y$, we have $r_i < R$, that is, the process choosing $r_i$ does not hit the cap of $R$ for every 
center in $Y$.
Note that, for a fixed iteration $i$, we have
$$\Pr(r_i = R) = \left(1-\frac{1}{k}\right)^{R-1} = \left(1-\frac{1}{k}\right)^{c_r k \log k - 1} \geq k^{-0.1 \cdot c_r}.$$
Thus, for sufficiently large constant $c_r$ (depending only on $\pconst$ and $\pdeg$), we have that 
$$\Pr(r_i = R) < \left(k \cdot |Y|\right)^{-1}.$$
We infer that, for such a choice of $c_r$, the probability that we are not lucky is at most $1/k$.

\textbf{Case 2: hitting a vertex within distance $r_i < R$.}
It is convenient to think here of the choice of the radius $r_i$ as an interative process
that starts from $r_i=1$, accepts the current radius with probability $1/k$, or increases its by one and repeats with probability $1-1/k$.
For a fixed iteration $i$ and a choice of $v_i$, consider a potential radius $r_i < R$ when there is a vertex $x \in X$ within distance exactly $r_i$ from $v_i$ in $G_{i-1}$.
If we do not accept this radius (which happens with probability $1-1/k$), the vertex $x$ is included in $\Ball_{G_{i-1}}(v_i,r_i)$ and is surely included in $G'$.
Consequently, in the whole process we care about not accepting a given radius only $k$ times, at most once for every vertex $x \in X$.
We infer that the probability that for some iteration $i$ there 
is a vertex $x \in X$ within distance exactly $r_i$ from $v_i$ and $r_i < R$
is at most $1-(1-1/k)^k$. 

Considering both cases, by union bound, the probability that $X \subseteq V(G')$ is at least
$$\left(1-\frac{1}{k}\right)^k - \frac{1}{k} \geq \frac{17}{256}.$$
The last estimate uses the assumption $k \geq 4$.
\cqed\end{proof}
Claim~\ref{cl:chop-hit} concludes the proof of Lemma~\ref{lem:alg1}.
\end{proof}

\subsection{Handling a component of polynomial size}

In this section we show the following lemma.
\begin{lemma}\label{lem:alg2}
Let $\mathcal{G}$ be as in Theorem~\ref{thm:alg}.
For every constant $c_r > 0$ there exists a constant $c > 0$ and a polynomial-time randomized algorithms that, given a positive integer $k$, and a connected graph $G \in \mathcal{G}$
of radius $c_r k \log k$, outputs a subset $A \subseteq V(G)$ such that
\begin{enumerate}
\item treedepth of $G[A]$ is $\Oh(k^{1 - \frac{1}{1+\pdeg}} \log k)$;
\item for every set $X \subseteq V(G)$ of size at most $k$, the probability that
$X \subseteq A$ is at least $2^{-c \cdot |X| \cdot k^{-\frac{1}{1+\pdeg}} \cdot \log^2 k}$.
\end{enumerate}
\end{lemma}
We emphasize here the linear dependency on $|X|$ in the exponent of the probability bound.
This dependency, similarly as in the analysis of~\cite{patcov-arxiv}, allows us to easily analyse independent runs of the algorithm on multiple connected components.

\begin{proof}[Proof of Lemma~\ref{lem:alg2}.]
The random process we employ is similar to the one of the previous section, but more involved.
Let $c_r' > 0$ be a constant to be fixed later.

We start with $G_0 = G$, $A_0 = \emptyset$ and $B_0 = \emptyset$. In the $i$-th iteration of the process, we consider the graph $G_{i-1}$.
If the graph $G_{i-1}$ is empty, we stop. Otherwise, we pick an arbitrary vertex $v_i \in V(G_{i-1})$ and pick a radius $r_i$
according to the geometric distribution with success probability $k^{-1/(1+\pdeg)} \log k$, capped at value $R' := c_r' k^{1/(1+\pdeg)}$
(i.e., as before, if the choice of the radius is greater than $R'$, we set $r_i := R'$).
 In other words, we start with $r_i=1$ and iteratively
accept the current radius with probability $k^{-1/(1+\pdeg)} \log k$ or increase it by one and repeat with the remaining probability, stopping unconditionally at radius $R'$.

As before, we set $A_i := A_{i-1} \cup \Ball_{G_{i-1}}(v_i, r_i)$
and $G_i := G_{i-1} - (\Ball_{G_{i-1}}(v_i, r_i) \cup \DBall_{G_{i-1}}(v_i, r_i))$. 
However, now, as the radii are smaller, we may want to retain some vertices of $\DBall_{G_{i-1}}(v_i, r_i)$, as they can be part of the pattern $X$; for this, we use the sets $B_i$.
With probability $1-1/(k|V(G)|)$ we put $P_i = \emptyset$ and $B_i = B_{i-1}$. 
With the remaining probability, we proceed as follows.
Uniformly at random, we choose a number $1 \leq \ell_i \leq k^{1-1/(1+\pdeg)} \log k$ and a set $P_i$ of $\ell_i$ vertices of $\DBall_{G_{i-1}}(v_i,r_i)$ (or all of them, if there are less than
$\ell_i$ vertices in this set). We put $B_i := B_{i-1} \cup P_i$.

Let $i_0$ be the index of the last iteration. If $|B_{i_0}| > k^{1-1/(1+\pdeg)} \log k$, the we output $A = \emptyset$.
Otherwise, we output $A := A_{i_0} \cup B_{i_0}$. Let us now verify that $A$ has the desired properties.

\begin{claim}\label{cl:poly:tw}
The treedepth of $G[A]$ is $\Oh(k^{\pdeg/(1+\pdeg)} \log k)$.
\end{claim}
\begin{proof}
The claim is trivial if $A = \emptyset$, so assume otherwise; in particular, $|B_{i_0}| \leq k^{1-1/(1+\pdeg)} \log k$.
We use the following inductive definition of treedepth: treedepth of an empty graph if $0$, while for any graph~$G$ on at least one vertex we have that
$$\mathrm{treedepth}(G) =
  \begin{cases}
    1 + \min\{\mathrm{treedepth}(G-v) : v \in V(G)\} & \textrm{if }G\textrm{ is connected} \\
    \max\{\mathrm{treedepth}(C) : C \textrm{ connected component of } G\} & \textrm{otherwise.} 
  \end{cases}$$
Upon deleting from $G[A]$ the at most $k^{1-1/(1+\pdeg)} \log k$ vertices of $B_{i_0}$, we are left with $G[A_{i_0}]$.
Similarly as in the previous section, every connected component of $G[A_{i_0}]$ is of radius at most $R' = c_r' k^{1/(1+\pdeg)}$.
Consequently, every connected component of $G[A_{i_0}]$ is of size at most $\pconst \cdot (c_r')^\pdeg k^{\pdeg/(1+\pdeg)}$.
The claim follows.
\cqed\end{proof}

\begin{claim}\label{cl:poly:hit}
For every set $X \subseteq V(G)$ of size at most $k$, the probability that $X \subseteq A$ is at least
$2^{-c |X| k^{-1/(1+\pdeg)} \log^2 k}$ for some constant $c>0$ depending only
on $c_r$, $\pdeg$, and $\pconst$.
\end{claim}
\begin{proof}
Fix a pattern $X$. The claim is trivial for $X = \emptyset$ so assume otherwise. In particular, if $|X| \geq 1$,
then we can estimate the desired probability as 
$$2^{-c |X| k^{-1/(1+\pdeg)} \log^2 k} = 1-\Omega\left(\frac{\log^2 k}{k^{1/(1+\pdeg)}}\right).$$

Consider a fixed iteration $i$, and the moment when, knowing $v_i$, we choose the radius $r_i$.
Given $G_{i-1}$ and $v_i$, we say that a radius $r$ is \emph{bad} if
\begin{equation}\label{eq:grow-ball}
\left|X \cap \DBall_{G_{i-1}}(v_i, r)\right| > \left(k^{-1/(1+\pdeg)} \log k\right)\cdot \left|X \cap \Ball_{G_{i-1}}(v_i,r)\right|,
\end{equation}
Let $1 \leq r^0 < r^1 < r^2 < \ldots r^t$ be a sequence of bad radii. First, note that 
$X \cap \DBall(v_i, r^0) \neq \emptyset$, and thus $|X \cap \Ball(v_i, r^1)| \geq 1$.
Furthermore, as for every $j \geq 1$ we have $\DBall(v_i,r^j) \subseteq \Ball(v_i, r^{j+1})$, we have
$$|X \cap \Ball(v_i, r^{j+1})| \geq \left(1+k^{-1/(1+\pdeg) \log k}\right) |X \cap \Ball(v_i,r^j)|.$$
Consequently,
  $$|X \cap \Ball(v_i, r^j)| \geq \left(1+k^{-1/(1+\pdeg) \log k}\right)^{j-1}.$$
Since $|X| \leq k$, we infer that
\begin{equation}\label{eq:t-bound}
t < 10 k^{1/(1+\pdeg)}.
\end{equation}

We are interested in the following event $\mathbf{A}$: every chosen radii $r_i$ is not bad and is smaller than $R'$ (i.e., we did not hit the cap of $R'$).
Recall the iterative interpretation of the choice of the radii $r_i$:
we start with $r_i=1$, accept the current radius with probability $k^{-1/(1+\pdeg)} \log k$, or increase $r_i$ by one and repeat with the remaining probability.
Thus, we are interested in the intersection of the following two events: we do not accept any bad radius, but we accept some good radius before the cap $R'$.

Whenever we do not accept a bad radius $r$, a vertex of $X \cap \DBall(v_i, r)$ is included in $\Ball(v_i,r_i) \subseteq A_i$. Consequently,
in the whole algorithm we encounter at most $|X|$ bad radii; each is indepently accepted with probability $k^{-1/(1+\pdeg)} \log k$.

By~\eqref{eq:t-bound}, in a fixed iteration $i$ there are at most $10 k^{1/(1+\pdeg)}$ bad radii. 
Consequently, if we count only acceptance of good radii, the probability that the radius $r_i$ reaches the bound $R'$ is at most
$$\left(1-k^{-1/(1+\pdeg)}\log k\right)^{(c_r'-10) k^{1/(1+\pdeg)}} \leq k^{-0.1c_r'}.$$
Consequently, since $|V(G)| \leq \pconst \cdot (c_r k \log k)^\pdeg$,
by choosing $c_r'$ large enough, we can ensure that the probability that there exists a radius $r_i$ equal to $R'$
is at most $k^{-1}$.
Since the choice of acceptance of different radii are independent, we infer that the probability of the event $\mathbf{A}$
is at least
$$\left(1-k^{-1}\right) \cdot \left(1-k^{-1/(1+\pdeg)} \log k\right)^{|X|} \geq 2^{-c_1 |X| k^{-1/(1+\pdeg)}\log k}$$
for some positive constant $c_1$. Here, we have used the fact that $|X| \geq 1$ and $k \geq 4$.

Assume that the event $\mathbf{A}$ happens, and let us fix one choice of $v_i$ and $r_i$. Note that these choices determine
the sets $A_i$ and the graphs $G_i$; the only remaining random choices are whether to include some vertices into the sets $B_i$.

For an iteration $i$, define $X_i := X \cap \DBall_{G_{i-1}}(v_i,r_i)$.
We are now considering the following event $\mathbf{B}$: in every iteration $i$ we have $P_i = X_i$. Note that if $\mathbf{B}$ happens, then $X \subseteq A$. Thus, we need to estimate
the probability of the event $\mathbf{B}$.

If $X_i = \emptyset$, then we guess so with probability $1-1/(k|V(G)|)$. As there are at most $|V(G)|$ iterations, 
with probability at least $1-1/k$ we will make correct decision in all iterations $i$ for which $X_i = \emptyset$.

Consider now an iteration $i$ for which $X_i \neq \emptyset$. Since the radius $r_i$ is good, we have
\begin{equation}\label{eq:grow-ball2}
\left|X \cap \DBall_{G_{i-1}}(v_i, r_i)\right| \leq k^{-1/(1+\pdeg)}\log k \left|X \cap \Ball_{G_{i-1}}(v_i,r_i)\right|.
\end{equation}
In particular, $|X \cap \Ball_{G_{i-1}}(v_i,r_i)| \geq k^{1/(1+\pdeg)}/\log k$, and thus there are at most
$k^{\pdeg/(1+\pdeg)} \log k$ such iterations. Furthermore,
$$\left|\bigcup_{i=1}^{i_0} X_i \right| \leq |X| k^{-1/(1+\pdeg)} \log k.$$
In every such iteration $i$, we need to correctly guess that $X_i$ is nonempty ($1/(k|V(G)|)$ success probability),
   correctly guess $\ell_i = |X_i|$ (at least $1/k$ success probability)
  and correctly guess $P_i = X_i$ (at least $|V(G)|^{-|X_i|}$ success probability). All these choices are independent.
  Since $|V(G)|$ is bounded polynomially in $k$, the probability of the event $\mathbf{B}$ is at least

\begin{align*}
\left(1-\frac{1}{k}\right)\cdot \prod_{i : X_i \neq \emptyset} \frac{1}{k|V(G)|} \cdot \frac{1}{k} \cdot \frac{1}{|V(G)|^{|X_i|}} &\geq \left(1-\frac{1}{k}\right)\cdot (|V(G)|^2 \cdot k)^{-|X| \cdot k^{-1/(1+\pdeg)}\log k} \\
    &\geq 2^{-c_2 |X| \cdot k^{-1/(1+\pdeg)} \log^2 k}
    \end{align*}
for some constant $c_2$ depending on $c_r$, $\pdeg$, and $\pconst$.
This finishes the proof of the claim.
\cqed\end{proof}
Lemma~\ref{lem:alg2} follows directly from Claims~\ref{cl:poly:tw} and~\ref{cl:poly:hit}.
\end{proof}

\subsection{Summary}

Let us now wrap up the proof of Theorem~\ref{thm:alg}, using Lemmata~\ref{lem:alg1} and~\ref{lem:alg2}.
We first apply the algorithm of Lemma~\ref{lem:alg1} to the input graph $G$ and integer $k$, obtaining a set $A_0 \subseteq V(G)$.
Then, we apply the algorithm of Lemma~\ref{lem:alg2} independently to every connected component $C$ of $G[A_0]$, obtaining a set $A_C \subseteq C$; recall that every such component
is of radius at most $R = c_r k \log k$. As the output $A$, we return the union of the returned sets $A_C$.
Clearly, the treedepth bound holds. If we denote $X_C := X \cap C$ for a component $C$, we have 
that the probability that $X \subseteq A$ is at least
$$\frac{17}{256} \cdot \prod_{C} 2^{-c |X_C| k^{-1/(1+\pdeg)} \log^2 k } \geq \frac{17}{256} \cdot 2^{-c k^{1-1/(1+\pdeg)} \log^2 k}.$$
This finishes the proof of Theorem~\ref{thm:alg}.

\section{Lower bound: proof of Theorem~\ref{thm:lb}}\label{sec:lower}

In this section we prove Theorem~\ref{thm:lb}.
The reduction is heavily inspired by the reduction for $\pdeg$-dimensional Euclidean TSP
by Marx and Sidiropolous~\cite{MarxS14}.
In particular, our starting point is the same CSP pivot problem.

\begin{theorem}[\cite{MarxS14}]\label{thm:csp}
For every fixed $\pdeg \geq 2$, there is a constant $\lambda_\pdeg$ 
such that for every constant $\varepsilon > 0$
an existence of an algorithm solving in time $2^{\Oh(n^{\pdeg-1-\varepsilon})}$ 
CSP instances with binary constraints, domain size at most $\lambda_\pdeg$, and Gaifman graph
being a $\pdeg$-dimensional grid of side length $n$ would refute ETH.
\end{theorem}

Let us recall that a \emph{binary CSP instance} consists of a \emph{domain} $D$, a set $V$ of \emph{variables},
and a set $E$ of \emph{constraints}. Every constraint is a binary relation $\psi_{u,v} \subseteq D \times D$ that binds two variables $u,v \in V$.
The goal is to find an assignment $\phi: V \to D$ that satisfies every constraint; a constraint $\psi_{u,v}$ is satisfied if $(\phi(u),\phi(v)) \in \psi_{u,v}$.
The \emph{Gaifman graph} of a binary CSP instance has vertex set $V$ and an edge $uv$ for every constraint $\psi_{u,v}$.

Similarly as in the case of~\cite{MarxS14}, our goal is to turn a given CSP instance
as in Theorem~\ref{thm:csp} and turn it into a Hamiltonian path instance by local gadgets.
That is, we are going to replace every variable of the CSP instance with a constant-size
gadget (i.e., with size depending only on $d$ and $\lambda_\pdeg$); the way the gadget is traversed
by the Hamiltonian path indicates the choice of the value of the variable. The neighboring
gadgets are wired up to ensure that the constraint binding them is satisfied.

\subsection{$2$-chains}
\begin{figure}[htbp]
\begin{center}
\includegraphics{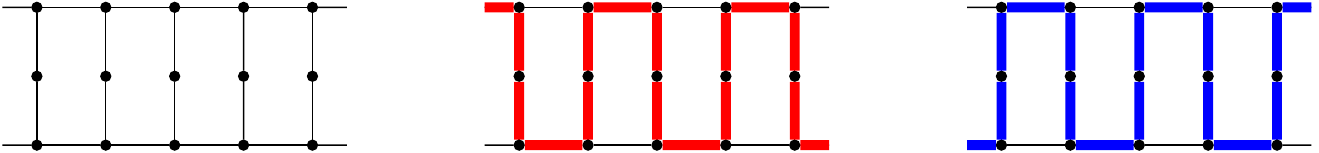}
\end{center}
\caption{A $2$-chain with two way how a Hamiltonian path can traverse it,
  called henceforth \emph{modes}.}\label{fig:red:2chain}
\end{figure}

The base gadget of the construction is a 2-chain as presented on Figure~\ref{fig:red:2chain}.
A direct check shows that there are two ways how a $2$-chain can be traversed by a Hamiltonian
path, as depicted on the figure.

\begin{figure}[htbp]
\begin{center}
\includegraphics{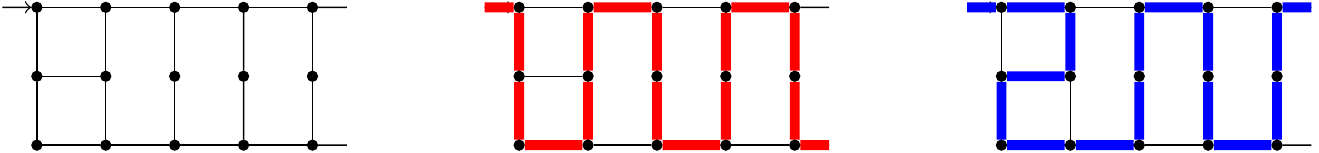}
\end{center}
\caption{An endpoint of a $2$-chain, allowing traversing the $2$-chain in both modes.}\label{fig:red:2chain-entry}
\end{figure}

Figure~\ref{fig:red:2chain-entry} shows a gadget present on both endpoints of a 2-chain. 
As shown on the figure, it allows choosing how the $2$-chain is traversed.

We will refer to the two depicted Hamiltonian paths of a $2$-chain as \emph{modes} of the chain.
Given one of the horizontal edges of the $2$-chain, a mode is \emph{consistent} with this edge
if the corresponding Hamiltonian path traverses the edge in question, and \emph{inconsistent}
otherwise.

\begin{figure}[htbp]
\begin{center}
\includegraphics{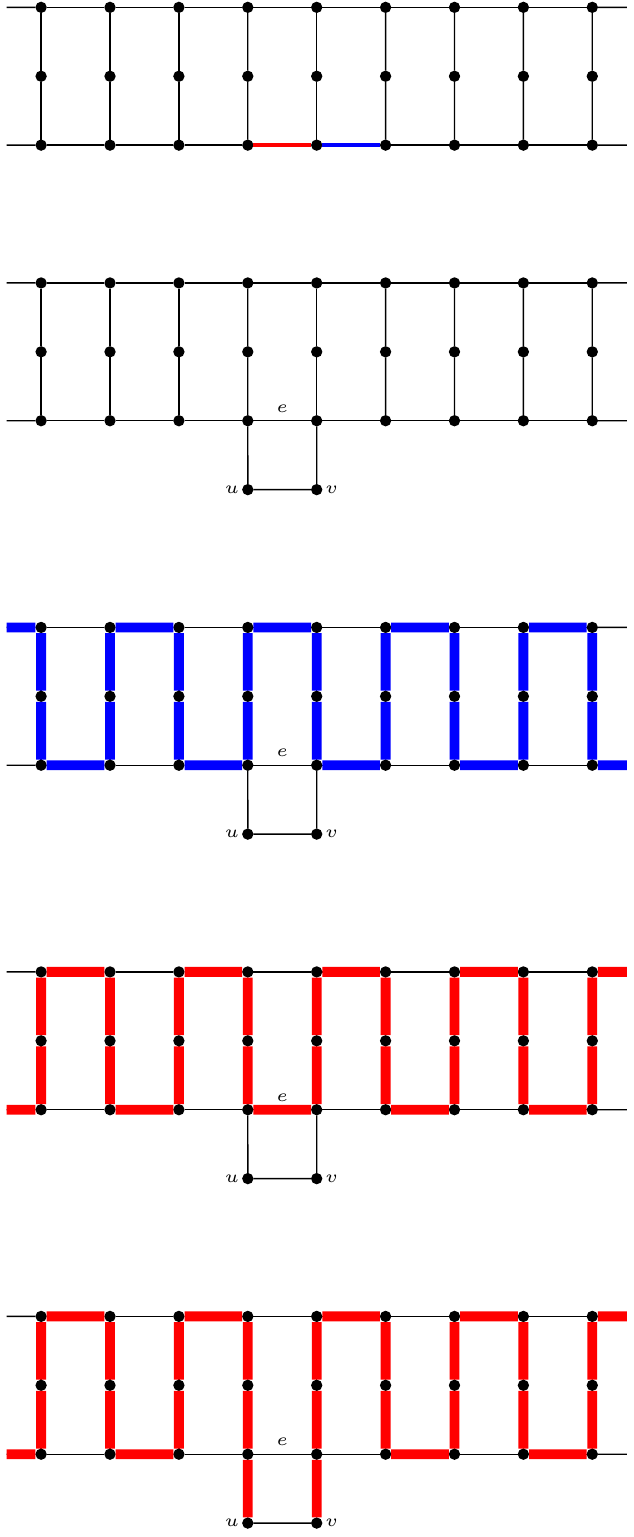}
\end{center}
\caption{From top to bottom: a chunk on a $2$-chain, with two attachment edges marked red and blue;
a standard attachment of a gadget; three ways how a $2$-chain with attached gadget can be traversed.}\label{fig:red:2chain-attach}
\end{figure}

We will attach various gadgets to $2$-chains via one of the horizontal edges.
To maintain the properties of the $2$-chains, in particular the effectively two ways 
of traversing a $2$-chain, we need to space out the attached gadgets. 
More formally, we partition every $2$-chain into sufficiently long chunks
(chunks of length $8$ are more than sufficient), and allow gadgets to attach only
to one of the two middle horizontal edges on one side of the chain
(see Figure~\ref{fig:red:2chain-attach}), with at most one gadget per chunk.
A gadget is always attached to an edge $e$ by adding two new vertices $u$ and $v$
near the edge $e$, in the same $2$-dimensional plane as the $2$-chain itself,
such that the endpoints of $e$, $u$, and $v$ form a square.
Properties of such an attachment can be summarized in the following straightforward claim.
\begin{claim}\label{cl:red:attach}
Consider a chunk $c$ on a $2$-chain $A$, and a gadget attached to an edge $e$ in $c$.
Then every Hamiltonian path traverses $c$ in one of the following three ways (see Figure~\ref{fig:red:2chain-attach}):
\begin{enumerate}
\item as on Figure~\ref{fig:red:2chain}, inconsistently with $e$;
\item as on Figure~\ref{fig:red:2chain}, consistently with $e$;
\item as on Figure~\ref{fig:red:2chain}, consistently with $e$, but with the edge $e$
replaced with an edge towards vertex $u$ and towards vertex $v$.
\end{enumerate}
\end{claim}
In particular, Claim~\ref{cl:red:attach} allows us to formally speak about a mode of a $2$-chain,
even if multiple gadgets are attached to it.

\subsection{Placing $2$-chains}
\begin{figure}[htbp]
\begin{center}
\includegraphics{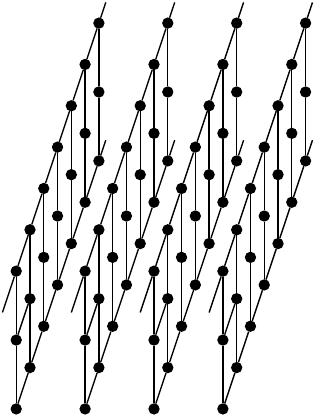}
\end{center}
\caption{Placing parallel $2$-chains for a single variable $x$.}\label{fig:red:2chain-placement}
\end{figure}

For every variable of the input CSP instance, we create $\lambda_\pdeg$ $2$-chains of length
$L = \Oh(d \lambda_\pdeg)$ (to be determined later). They are positioned parallely in the following fashion (see Figure~\ref{fig:red:2chain-placement}):
we choose an arbitrary $3$-dimensional subspace
of the whole grid, and place $2$-chains such that $i$-th $2$-chain occupies
vertices $\{0,1,\ldots,L\} \times \{0,1,2\} \times \{i\}$.
The edges indicated as attachment points for gadgets are on the one side of all chains.

All chains, for all variables, are wired up into a Hamiltonian path: for every variable,
we connect the constructed $2$-chains into a path in a straightforward fashion, we take an arbitrary
Hamiltonian path of the original Gaifman graph of the input CSP instance (which is a $\pdeg$-dimensional grid, and thus trivially admits a Hamiltonian path),
and connect endpoints
of the $2$-chains in the same order using simple paths. This is straightforward to perform
if we space out the variable gadgets enough.

Since all constructed $2$-chains are isomorphic, we indicate one mode of a $2$-chain as a \emph{low mode}, and the other one as \emph{high mode}. Our goal is to introduce gadgets that (i) ensure that for every variable, exactly one of the corresponding $2$-chains is in high mode, indicating
the choice of the value for this variable; (ii) for every two variables that are bound by a
constraint, for every pair of values that is forbidden by the constraint, ensure that the
two variables in question do not attain the values in quesion at the same time, that is,
the corresponding two $2$-chains are not both in high mode at the same time.

\subsection{OR-checks}
The construction of $2$-chains allow us to implement a simple ``OR'' constraint on two $2$-chains.
Consider two $2$-chains $A$ and $B$, and two horizontal edges $e_A$ and $e_B$ on $A$ and $B$, respectively.
By attaching an OR-check to these edges we mean the following construction:
\begin{enumerate}
\item we create vertices $u_A$ and $v_A$ near $e_A$ as well as $u_B$ and $v_B$ near $e_B$, as in the description of gadget attachment;
\item we connect $u_A$ to $u_B$ by a path and $v_A$ to $v_B$ by a path.
\end{enumerate}
If the $2$-chains are spaced enough, it is straightforward to implement the above constuction 
such that the resulting graph is a subgraph of a $d$-dimensional grid.

Claim~\ref{cl:red:attach} allows us to observe the following.
\begin{claim}\label{cl:red:OR}
If $A$ is traversed in a way consistent with $e_A$, then one can modify the Hamiltonian path
traversing $A$ so that it visits the OR gadget: replace $e_A$ with a path traversing first a path from $u_A$ to $u_B$, the edge $u_Bv_B$, and then the path from $v_B$ to $v_A$.
A symmetrical claim holds if $B$ is traversed in a way consistent with $e_B$.

In the other direction, there is no Hamiltonian path that traverses both $A$ and $B$
in a way inconsistent with $e_A$ and $e_B$, respectively.
\end{claim}

We now observe that, by attaching OR-checks in a straightforward manner, we can ensure that:
\begin{enumerate}
\item for every variable $x$, at most one $2$-chain corresponding to $x$ is in high mode
(we wire up every pair of $2$-chains with an OR-check forbidding two high modes at the same time);
\item for every two variables $x$ and $y$ that are bound by a constraint $\psi$, 
  for every pair of values $(\alpha_x,\alpha_y)$ that is forbidden by the constraint $\psi$,
  the $\alpha_x$-th $2$-chain of $x$ and the $\alpha_y$-th $2$-chain of $y$ are not in the
  high mode at the same time.
\end{enumerate}
We are left with ensuring that for every variable $x$, at least one of the corresponding
$2$-chains is in the high mode. This is the aim of the next gadget.

\subsection{Tube gadget}
\begin{figure}[htbp]
\begin{center}
\includegraphics{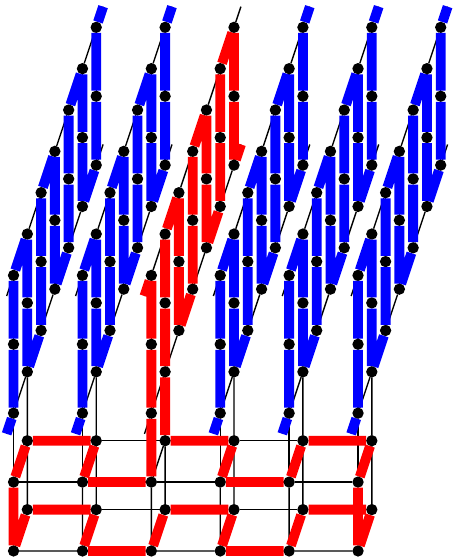}
\end{center}
\caption{A tube gadgets attached to the $2$-chains, with intendent Hamiltonian path.}\label{fig:red:tube}
\end{figure}

Fix a variable $x$. Without loss of generality, we can assume that the first chunk
of every $2$-chain for $x$ has not been used by the OR-checks introduced previously.
Let $e_i$ be the attachment edge of the $i$-th $2$-chain that is consistent with the high
mode of the $2$-chain; note that the edges $e_i$ lie next to each other (see Figure~\ref{fig:red:tube}).

We create a $2 \times 2 \times \lambda_\pdeg$ grid, called henceforth a \emph{tube gadget}, placed near the edges $e_i$, such that
every edge $e_i$ can be attached to an edge of the grid in a standard way discussed earlier.
See Figure~\ref{fig:red:tube} for an illustration.

Since a $2 \times 2 \times \lambda_\pdeg$ grid admits a Hamiltonian cycle that traverses
every edge in one of the ``short'' directions, if the $i$-th chain is traversed
in high mode for some $i$, we can replace $e_i$ on the Hamiltonian path with a traverse
along the aforementioned Hamiltonian cycle. This observation, together
with Claim~\ref{cl:red:attach}, proves the following claim.
\begin{claim}\label{cl:red:tube}
If there exists an index $i$ such that the $i$-th $2$-chain is traversed in high mode,
   then the Hamiltonian path of this $2$-chain can be altered to visit every vertex
   of the $2 \times 2 \times \lambda_\pdeg$ grid.

On the other hand, any Hamiltonian path of the entire graph needs to traverse at least
one $2$-chain in high mode, in order to visit the vertices of the $2 \times 2 \times \lambda_\pdeg$ grid.
\end{claim}

\subsection{Summary}
The tube gadgets ensure that, for every variable, at least one corresponding $2$-chain
is in high mode. The first type of the attached OR-checks ensure that at most one such $2$-chain
is in high mode. Thus, effectively the gadgets introduced for a single variable $x$
can be in one of $\lambda_\pdeg$ by choosing the $2$-chain that is in high mode, which
corresponds to the choice of the value for $x$ in an assignment.

The second type of the attached OR-checks ensure that the values of the neighboring
variables satisfy the constraint that binds them, completing the proof of the correctness
of the reduction.

To conclude, let us observe that every $2$-chain is attached to one tube gadget
and $\Oh(d \lambda_\pdeg)$ OR-checks, and the whole gadget replacing a single variable takes part in $\Oh(d \lambda_\pdeg^2)$ OR-checks.
Thus taking $L = \Oh(d \lambda_\pdeg^2)$ suffices.
By leaving space of size $\Oh(d \lambda_\pdeg^2)$ between consecutive variable gadgets
we can ensure more than enough space for all connections. Consequently,
   the constructed graph is a subgraph of a $d$-dimensional grid of side length
   $\Oh(d \lambda_\pdeg^2 n)$, and admits a Hamiltonian path if and only if the input
   CSP instance is satisfiable. This finishes the proof of Theorem~\ref{thm:lb}.

\section{Conclusions}\label{sec:conc}

We have shown a low treewidth pattern covering statement for graphs of polynomial growth
with subexponential term being $2^{k^{1-\frac{1}{1+\pdeg}}}$, where $\pdeg$ is the growth rate of
the graph. An almost tight lower bound shows that, assuming ETH, one should not hope
for a better term than $2^{k^{1-\frac{1}{\pdeg}}}$.

Two natural questions arise. The first one is to close the gap between $\frac{1}{1+\pdeg}$
and $\frac{1}{\pdeg}$; we conjecture that our lower bound is tight, and the term
$k^{1-\frac{1}{1+\pdeg}}$ in the running time bound of Theorem~\ref{thm:alg} is only
a shortfall of our algorithmic techniques. 
The second one is to derandomize the algorithms of this work and of~\cite{patcov-arxiv,patcov-focs}.
The clustering step is the only step of the algorithm of~\cite{patcov-arxiv,patcov-focs}
that we do not now how to derandomize, despite its resemblance to the construction
of Bartal's HSTs~\cite{Bartal98} that were subsequently derandomized~\cite{CharikarCGGP98}.

\newpage

\bibliographystyle{abbrv}
\bibliography{pat-cov,book_pc}

\begin{thebibliography}{10}

\bibitem{AbrahamGGM06}
I.~Abraham, C.~Gavoille, A.~V. Goldberg, and D.~Malkhi.
\newblock Routing in networks with low doubling dimension.
\newblock In {\em 26th {IEEE} International Conference on Distributed Computing
  Systems {(ICDCS} 2006), 4-7 July 2006, Lisboa, Portugal}, page~75. {IEEE}
  Computer Society, 2006.

\bibitem{DBLP:conf/spaa/AbrahamM05}
I.~Abraham and D.~Malkhi.
\newblock Name independent routing for growth bounded networks.
\newblock In P.~B. Gibbons and P.~G. Spirakis, editors, {\em {SPAA} 2005:
  Proceedings of the 17th Annual {ACM} Symposium on Parallelism in Algorithms
  and Architectures, July 18-20, 2005, Las Vegas, Nevada, {USA}}, pages 49--55.
  {ACM}, 2005.

\bibitem{Bartal98}
Y.~Bartal.
\newblock On approximating arbitrary metrices by tree metrics.
\newblock In J.~S. Vitter, editor, {\em Proceedings of the Thirtieth Annual
  {ACM} Symposium on the Theory of Computing, Dallas, Texas, USA, May 23-26,
  1998}, pages 161--168. {ACM}, 1998.

\bibitem{BlondelJKS13}
V.~Blondel, K.~Jung, P.~Kohli, and D.~Shah.
\newblock Partition-merge: Distributed inference and modularity optimization.
\newblock {\em CoRR}, abs/1309.6129, 2013.

\bibitem{chan-thesis}
T.~H.~H. Chan.
\newblock {\em Approximation Algorithms for Bounded Dimensional Metric Spaces}.
\newblock PhD thesis, Carnagie Mellon University, 2007.
\newblock Available at \texttt{http://i.cs.hku.hk/~hubert/thesis/thesis.pdf}.

\bibitem{CharikarCGGP98}
M.~Charikar, C.~Chekuri, A.~Goel, S.~Guha, and S.~A. Plotkin.
\newblock Approximating a finite metric by a small number of tree metrics.
\newblock In {\em 39th Annual Symposium on Foundations of Computer Science,
  {FOCS} '98, November 8-11, 1998, Palo Alto, California, {USA}}, pages
  379--388. {IEEE} Computer Society, 1998.

\bibitem{DBLP:conf/soda/ChitnisHM14}
R.~H. Chitnis, M.~Hajiaghayi, and D.~Marx.
\newblock Tight bounds for {P}lanar {S}trongly {C}onnected {S}teiner {S}ubgraph
  with fixed number of terminals (and extensions).
\newblock In {\em SODA 2014}, pages 1782--1801, 2014.

\bibitem{thebook}
M.~Cygan, F.~V. Fomin, L.~Kowalik, D.~Lokshtanov, D.~Marx, M.~Pilipczuk,
  M.~Pilipczuk, and S.~Saurabh.
\newblock {\em Parameterized Algorithms}.
\newblock Springer, 2015.

\bibitem{DBLP:journals/siamdm/DemaineFHT04}
E.~D. Demaine, F.~V. Fomin, M.~T. Hajiaghayi, and D.~M. Thilikos.
\newblock Bidimensional parameters and local treewidth.
\newblock {\em SIAM J. Discrete Math.}, 18(3):501--511, 2004.

\bibitem{DBLP:journals/talg/DemaineFHT05}
E.~D. Demaine, F.~V. Fomin, M.~T. Hajiaghayi, and D.~M. Thilikos.
\newblock Fixed-parameter algorithms for $(k,r)$-{C}enter in planar graphs and
  map graphs.
\newblock {\em ACM Transactions on Algorithms}, 1(1):33--47, 2005.

\bibitem{DemaineFHT05}
E.~D. Demaine, F.~V. Fomin, M.~T. Hajiaghayi, and D.~M. Thilikos.
\newblock Subexponential parameterized algorithms on bounded-genus graphs and
  {$H$}-minor-free graphs.
\newblock {\em J. {ACM}}, 52(6):866--893, 2005.

\bibitem{DBLP:journals/cj/DemaineH08}
E.~D. Demaine and M.~Hajiaghayi.
\newblock The bidimensionality theory and its algorithmic applications.
\newblock {\em Comput. J.}, 51(3):292--302, 2008.

\bibitem{DBLP:journals/combinatorica/DemaineH08}
E.~D. Demaine and M.~Hajiaghayi.
\newblock Linearity of grid minors in treewidth with applications through
  bidimensionality.
\newblock {\em Combinatorica}, 28(1):19--36, 2008.

\bibitem{DBLP:conf/gd/DemaineH04}
E.~D. Demaine and M.~T. Hajiaghayi.
\newblock Fast algorithms for hard graph problems: Bidimensionality, minors,
  and local treewidth.
\newblock In {\em Graph Drawing}, pages 517--533, 2004.

\bibitem{DBLP:conf/stacs/DornFLRS10}
F.~Dorn, F.~V. Fomin, D.~Lokshtanov, V.~Raman, and S.~Saurabh.
\newblock Beyond bidimensionality: {P}arameterized subexponential algorithms on
  directed graphs.
\newblock In {\em STACS 2010}, pages 251--262, 2010.

\bibitem{DBLP:journals/csr/DornFT08}
F.~Dorn, F.~V. Fomin, and D.~M. Thilikos.
\newblock Subexponential parameterized algorithms.
\newblock {\em Computer Science Review}, 2(1):29--39, 2008.

\bibitem{DornPBF10}
F.~Dorn, E.~Penninkx, H.~L. Bodlaender, and F.~V. Fomin.
\newblock Efficient exact algorithms on planar graphs: Exploiting sphere cut
  decompositions.
\newblock {\em Algorithmica}, 58(3):790--810, 2010.

\bibitem{patcov-focs}
F.~V. Fomin, D.~Lokshtanov, D.~Marx, M.~Pilipczuk, M.~Pilipczuk, and
  S.~Saurabh.
\newblock Subexponential parameterized algorithms for planar and
  apex-minor-free graphs via low treewidth pattern covering.
\newblock In {\em FOCS}, 2016.
\newblock To appear.

\bibitem{patcov-arxiv}
F.~V. Fomin, D.~Lokshtanov, D.~Marx, M.~Pilipczuk, M.~Pilipczuk, and
  S.~Saurabh.
\newblock Subexponential parameterized algorithms for planar and
  apex-minor-free graphs via low treewidth pattern covering.
\newblock {\em CoRR}, abs/1604.05999, 2016.

\bibitem{DBLP:journals/ipl/FominLRS11}
F.~V. Fomin, D.~Lokshtanov, V.~Raman, and S.~Saurabh.
\newblock Subexponential algorithms for partial cover problems.
\newblock {\em Inf. Process. Lett.}, 111(16):814--818, 2011.

\bibitem{DBLP:journals/siamcomp/FominT06}
F.~V. Fomin and D.~M. Thilikos.
\newblock Dominating sets in planar graphs: Branch-width and exponential
  speed-up.
\newblock {\em SIAM J. Comput.}, 36(2):281--309, 2006.

\bibitem{DBLP:conf/infocom/GummadiJSS09}
R.~Gummadi, K.~Jung, D.~Shah, and R.~S. Sreenivas.
\newblock Computing the capacity region of a wireless network.
\newblock In {\em {INFOCOM} 2009. 28th {IEEE} International Conference on
  Computer Communications, Joint Conference of the {IEEE} Computer and
  Communications Societies, 19-25 April 2009, Rio de Janeiro, Brazil}, pages
  1341--1349. {IEEE}, 2009.

\bibitem{eth}
R.~Impagliazzo, R.~Paturi, and F.~Zane.
\newblock Which problems have strongly exponential complexity?
\newblock {\em J. Comput. Syst. Sci.}, 63(4):512--530, 2001.

\bibitem{DBLP:conf/icalp/KleinM12}
P.~N. Klein and D.~Marx.
\newblock Solving planar $k$-terminal cut in ${O}(n^{c \sqrt{k}})$ time.
\newblock In {\em Proceedings of the 39th International Colloquium of Automata,
  Languages and Programming (ICALP)}, volume 7391 of {\em Lecture Notes in
  Comput. Sci.}, pages 569--580. Springer, 2012.

\bibitem{DBLP:conf/soda/KleinM14}
P.~N. Klein and D.~Marx.
\newblock A subexponential parameterized algorithm for {S}ubset {TSP} on planar
  graphs.
\newblock In {\em SODA 2014}, pages 1812--1830, 2014.

\bibitem{LinialS93}
N.~Linial and M.~E. Saks.
\newblock Low diameter graph decompositions.
\newblock {\em Combinatorica}, 13(4):441--454, 1993.

\bibitem{eth-surveye}
D.~Lokshtanov, D.~Marx, and S.~Saurabh.
\newblock Lower bounds based on the exponential time hypothesis.
\newblock {\em Bulletin of the {EATCS}}, 105:41--72, 2011.

\bibitem{MarxS14}
D.~Marx and A.~Sidiropoulos.
\newblock The limited blessing of low dimensionality: when 1-1/d is the best
  possible exponent for d-dimensional geometric problems.
\newblock In S.~Cheng and O.~Devillers, editors, {\em 30th Annual Symposium on
  Computational Geometry, SOCG'14, Kyoto, Japan, June 08 - 11, 2014}, page~67.
  {ACM}, 2014.

\bibitem{sparsity}
J.~Ne\v{s}et\v{r}il and P.~O. de~Mendez.
\newblock {\em Sparsity - Graphs, Structures, and Algorithms}, volume~28 of
  {\em Algorithms and combinatorics}.
\newblock Springer, 2012.

\bibitem{DBLP:conf/stacs/PilipczukPSL13}
M.~Pilipczuk, M.~Pilipczuk, P.~Sankowski, and E.~J. van Leeuwen.
\newblock Subexponential-time parameterized algorithm for {S}teiner {T}ree on
  planar graphs.
\newblock In {\em STACS 2013}, pages 353--364, 2013.

\bibitem{DBLP:conf/focs/PilipczukPSL14}
M.~Pilipczuk, M.~Pilipczuk, P.~Sankowski, and E.~J. van Leeuwen.
\newblock Network sparsification for {S}teiner problems on planar and
  bounded-genus graphs.
\newblock In {\em FOCS 2014}, pages 276--285. {IEEE} Computer Society, 2014.

\bibitem{DBLP:conf/esa/Thilikos11}
D.~M. Thilikos.
\newblock Fast sub-exponential algorithms and compactness in planar graphs.
\newblock In {\em ESA 2011}, pages 358--369, 2011.

\end{thebibliography}

\end{document}